

\input harvmac

\overfullrule=0pt


\def\L{{\scriptscriptstyle L}}

\def\P{{\scriptscriptstyle P}}
\def\Q{{\scriptscriptstyle Q}}
\def\R{{\scriptscriptstyle R}}
\def\S{{\scriptscriptstyle S}}
\def\T{{\scriptscriptstyle T}}


\def\CD{{\cal D}}

\def\CL{{\cal L}}

\def\CQ{{\cal Q}}
\def\CR{{\cal R}}
\def\CS{{\cal S}}
\def\CT{{\cal T}}


\def\d{\delta}
\def\e{\epsilon}
\def\g{\gamma}
\def\l{\lambda}

\def\o{\sigma}

\def\u{\mu}
\def\v{\nu}


\def\aEM{\alpha_{\scriptscriptstyle EM}}

\def\bar#1{\overline{#1}}
\def\ccdot{\hbox{\kern-.1em$\cdot$\kern-.1em}}
\def\cR{c_\R}
\def\cRS{c_{\R\S}}
\def\cRT{c_{\R\T}}
\def\cS{c_\S}
\def\cST{c_{\S\T}}
\def\dash{{\> \over \,\>}} 		

\def\EM{{\scriptscriptstyle EM}}

\def\gfive{\gamma^5}
\def\gtap{\raise.3ex\hbox{$>$\kern-.75em\lower1ex\hbox{$\sim$}}}

\def\ltap{\raise.3ex\hbox{$<$\kern-.75em\lower1ex\hbox{$\sim$}}}
\def\LX{{\Lambda_\chi}}

\def\mc{m_c}

\def\MeV{\,\> {\rm MeV}}
\def\mpi{m_{\pi}}

\def\mQ{m_\Q}

\def\proj{{1 + \slash{v} \over 2}}

\def\slash#1{#1\hskip-0.5em /}
\def\space{\>\>}
\def\sqrroot{\sqrt{(E_1^2-\mpi^2)(E_2^2-\mpi^2)}}


\def\half{{1 \over 2}}

\def\sixth{{ 1\over 6}}
\def\third{{1 \over 3}}
\def\threehalves{{3 \over 2}}

\def\twothirds{{2 \over 3}}


\newdimen\pmboffset
\pmboffset 0.022em
\def\oldpmb#1{\setbox0=\hbox{#1}%
 \copy0\kern-\wd0
 \kern\pmboffset\raise 1.732\pmboffset\copy0\kern-\wd0
 \kern\pmboffset\box0}
\def\pmb#1{\mathchoice{\oldpmb{$\displaystyle#1$}}{\oldpmb{$\textstyle#1$}}
      {\oldpmb{$\scriptstyle#1$}}{\oldpmb{$\scriptscriptstyle#1$}}}

\def\pib{{\pmb{\pi}}}


%
%
\def\appendix#1#2{\global\meqno=1\global\subsecno=0\xdef\secsym{\hbox{#1.}}
\bigbreak\bigskip\noindent{\bf Appendix. #2}\message{(#1. #2)}
\writetoca{Appendix {#1.} {#2}}\par\nobreak\medskip\nobreak}


\nfig\spectrum{Lowest lying $I=0$ and $I=1$ charmed baryon states.
Experimentally measured mass splittings in MeV of the baryons above the
$\Lambda_c (2286 \MeV)$ ground state are indicated in parentheses.
The dominant pion decay modes of the excited $J^\P=\half^-$
$\Lambda_{c1}$ and $J^\P=\threehalves^-$ $\Lambda_{c1}^*$ states are
illustrated by the solid and dashed lines respectively.  Their allowed
radiative transitions down to the ground state are represented by the
squiggly curves.}
\nfig\polegraphs{Leading order pole graphs that contribute to
$\Lambda_{c1}^{(*)} \to \Lambda_c \pi \pi$.}
\nfig\diffrate{Dimensionless differential decay rates
$h_2^{-2} d\Gamma(\Lambda_{c1} \to \Lambda_c \pi^0 \pi^0)/dE_1$ (solid curve)
and $10 \times h_2^{-2} d\Gamma(\Lambda_{c1}^* \to \Lambda_c \pi^0 \pi^0)/dE_1$
(dashed curve) plotted against pion energy $E_1$ measured in the excited
charm baryon's rest frame.  The coupling constant $h_1$ is set equal to
unity in this graph.}
\nfig\twopirate{Integrated double pion decay rates of $\Lambda_{c1}$
(solid curve) and $\Lambda_{c1}^*$ (dashed curve) plotted as functions of
coupling constant $h_1$.  The neutral and charged pion channel contributions
are summed together in this graph.}
%


\def\CITTitle#1#2#3{\nopagenumbers\abstractfont
\hsize=\hstitle\rightline{#1}
\vskip 0.6in\centerline{\titlefont #2} \centerline{\titlefont #3}
\abstractfont\vskip .5in\pageno=0}
\CITTitle{{\baselineskip=12pt plus 1pt minus 1pt
  \vbox{\hbox{CALT-68-1912}\hbox{DOE RESEARCH AND}\hbox{DEVELOPMENT
  REPORT}}}} {Strong and Electromagnetic Decays}{of Two New $\Lambda_c^*$
  Baryons}
\centerline{
  Peter Cho\footnote{$^\dagger$}{Work supported in part by an
  SSC Fellowship and by the U.S. Dept. of Energy under DOE Grant no.
  DE-FG03-92-ER40701.}}

\centerline{Lauritsen Laboratory}
\centerline{California Institute of Technology}
\centerline{Pasadena, CA  91125}

\vskip .3in
\centerline{\bf Abstract}
\bigskip

	Two recently discovered excited charm baryons are studied within
the framework of Heavy Hadron Chiral Perturbation Theory.  We interpret these
new baryons which lie $308 \MeV$ and $340 \MeV$ above the $\Lambda_c$
as $I=0$ members of a P-wave spin doublet.  Differential and total decay rates
for their double pion transitions down to the $\Lambda_c$ ground state are
calculated.  Estimates for their radiative decay rates are also discussed.  We
find that the experimentally determined characteristics of the
$\Lambda_c^*$ baryons may be simply understood in the effective theory.

\Date{1/94}


\nref\ARGUS{H. Albrecht {\it et al.} (ARGUS Collaboration), Phys. Lett.
 {\bf  B317} (1993) 227.}
\nref\CLEOI{D. Acosta {\it et al.} (CLEO Collaboration), ``Observation of
 the Excited Charmed Baryon $\Lambda_c^*$ in CLEO-II'', CLEO CONF 93-7,
 contributed to the International Symposium on Lepton and Photon Interactions,
 Ithaca, 1993.}
\nref\Fermigrp{P.L. Frabetti {\it et al.} (E687 Collaboration),
 FERMILAB-Pub-93-323-E (1993).}
\nref\CLEOII{M. Battle {\it et al.} (CLEO Collaboration), ``Observation of a
 New Excited Charmed Baryon'', CLEO CONF 93-32, contributed to the
 International Symposium on Lepton and Photon Interactions, Ithaca, 1993.}
\nref\WiseI{M. Wise, Phys. Rev. {\bf D45} (1992) R2188.}
\nref\Burdman{G. Burdman and J. Donoghue, Phys. Lett. {\bf B280} (1992) 287.}
\nref\Yan{H.Y. Cheng, C.Y. Cheung, G.L. Lin, Y.C. Lin, T.M. Yan and H.L.
  Yu, Phys. Rev. {\bf D46} (1992) 1148.}
\nref\ChoI{P. Cho, Phys. Lett. {\bf B285} (1992) 145.}
\nref\WiseII{For a general review of Heavy Hadron Chiral Perturbation
 Theory, see M. Wise, Caltech Preprint CALT-68-1860 (1993), Lectures given
 at the CCAST Symposium.}
\nref\FalkLuke{A. Falk and M. Luke, Phys. Lett. {\bf B292} (1992) 119.}
\nref\Kilian{U. Kilian, J.G. K\"orner and D. Pirjol, Phys. Lett. {\bf B288}
 (1992) 360.}
\nref\ChoTrivedi{P. Cho and S. Trivedi, Caltech preprint CALT-68-1902 (1993).}
\nref\Copley{L.A. Copley, N. Isgur and G. Karl, Phys. Rev. {\bf D20} (1979)
 768.}
\nref\Georgi{H. Georgi, Nucl. Phys. {\bf B348} (1991) 293.}
\nref\LukeManohar{M. Luke and A. Manohar, Phys. Lett. {\bf B286} (1992)
 348.}
\nref\ChoII{P. Cho, Nucl. Phys. {\bf B396} (1993) 183; Caltech preprint
 CALT-68-1884.}
\nref\NDA{
  H. Georgi, Phys. Lett. {\bf B298} (1993) 187\semi
  A. Manohar and H. Georgi, Nucl. Phys. {\bf B234} (1984) 189\semi
  H. Georgi and L. Randall, Nucl. Phys. {\bf B276} (1986) 241.}
\nref\Appel{J.A. Appel, FERMILAB-Conf-93/328 (1993).}
\nref\SKAT{V.V. Ammosov {\it et al.} (SKAT Collaboration), ``Observation of
 $\Sigma_c^{*++}$ production in neutrino interactions on bubble chamber
 SKAT'', paper contributed to the International Symposium on Lepton and Photon
 Interactions, Ithaca, 1993.}
\nref\Sigmamasses{G. Crawford {\it et al.} (CLEO Collaboration), CLEO
 Preprint CLNS 93/1230 (1993).}
\nref\bounce{M. Wise, private communication.}
\nref\ChoGeorgi{P. Cho and H. Georgi, Phys. Lett. {\bf B296} (1992) 408;
  (E) Phys. Lett. {\bf B300} (1993) 410.}
\nref\Amundson{J.F. Amundson, C.G. Boyd, E. Jenkins, M. Luke, A.V. Manohar,
 J. Rosner, M. Savage and M.B. Wise, Phys. Lett. {\bf B296} (1992) 415.}
\nref\YanII{H.Y. Cheng, C.Y. Cheung, G.L. Lin, Y.C. Lin, T.M. Yan and H.L.
 Yu, Phys. Rev. {\bf D47} (1992) 1030.}
\nref\YanIII{H.Y. Cheng, C.Y. Cheung, G.L. Lin, Y.C. Lin, T.M. Yan and H.L.
 Yu, Cornell preprint CLNS 93/1189 (1993).}
\nref\IsgurWise{N. Isgur and M. Wise, Phys. Rev. Lett. {\bf 66} (1991)
 1130.}

\newsec{Introduction}

	The discovery of the first excited charm baryon has recently been
announced by the ARGUS, CLEO and E687 groups \refs{\ARGUS{--}\Fermigrp}.
The new state lies approximately $340 \MeV$ above the $\Lambda_c
(2286 \MeV)$ and decays to it via double pion emission.
Although its spin, isospin and parity are not yet known, this new
charmed baryon has been preliminarily interpreted as a $\Lambda_c^*$
resonance.  CLEO has further reported evidence for a second $\Lambda_c^*$
excitation at $308 \MeV$ above $\Lambda_c$ \CLEOII.  The second resonance
also decays through a double pion mode that is consistent with the two step
process $\Lambda_c^* \to \Sigma_c \pi$ followed by
$\Sigma_c \to \Lambda_c \pi$.  In contrast, CLEO finds no evidence for an
intermediate $\Sigma_c$ in the decay of the first $\Lambda_c^*$ excitation
\CLEOI.

	In this article, we will analyze these new baryon states and their
dominant decay modes within the framework of Heavy Hadron Chiral Perturbation
Theory (HHCPT).  This hybrid effective theory represents a synthesis of Chiral
Perturbation Theory and the Heavy Quark Effective Theory (HQET) and describes
the low energy interactions between light Goldstone bosons and hadrons
containing a heavy quark \refs{\WiseI{--}\WiseII}.   Since its development a
few years ago, HHCPT has primarily been applied to the study of ground
state charm and bottom hadrons.   Ground state mesons and baryons are more
tightly restricted by heavy quark spin symmetry than their excited
counterparts.  Moreover, experimental information has been much more sparse
for the latter than the former.  It is therefore not surprising that
theorists have concentrated upon the lowest lying hadrons in
the past.   Now however that new data is being collected, it is worthwhile
to broaden the scope of HHCPT and incorporate excited heavy hadrons into the
effective theory.

	The first excited heavy mesons and baryons are P-wave hadrons
that carry one unit of orbital angular momentum.  P-wave mesons have already
been investigated within the HHCPT framework \refs{\FalkLuke{--}\ChoTrivedi}.
It is straightforward to extend the formalism and include P-wave baryons
as well.  A number of unknown couplings enter into the excited baryon sector
which limits one's predictive power.   But as we shall see, all the
general characteristics of the two $\Lambda_c^*$ baryons reported by
ARGUS, CLEO and E687 are consistent with their being members of an excited
spin symmetry doublet.  Although our findings will be more qualitative
than quantitative, we hope this work may help guide experimentalists as
they continue to study these new charmed baryons.

	Our paper is organized as follows.  In section~2, we incorporate
the lowest lying excited baryon doublet into the heavy baryon chiral
Lagrangian.  We then focus upon the two new $\Lambda_c^*$ members of this
doublet and analyze their strong interaction decays in section~3.
Radiative transitions are discussed in section~4.  Finally, we
close in section~5 with some thoughts on future directions for
investigation.

\newsec{The Heavy Baryon Chiral Lagrangian}

	We begin by recalling some basic aspects of the baryon sector in
Heavy Hadron Chiral Perturbation Theory \refs{\Yan,\ChoI}.  Ground state
baryons with quark content $Qqq$ have zero orbital angular momentum and
occur in two types depending upon the angular momentum $j_\ell$ of their
light degrees of freedom.  In the first case, the light brown muck is
arranged in a symmetric $j_\ell=1$ configuration which transforms as a
sextet under flavor $SU(3)$.  The spectators consequently couple with the
heavy quark to form $J^\P = \half^+$ and
$J^\P = \threehalves^+$ S-wave bound states.  When the heavy quark is taken
to be charm, the spin-$\half$ states are annihilated by velocity
dependent Dirac operators $S^{ij}(v)$ whose individual components are
given by
\eqn\sextetcomp{\eqalign{
 S^{11} &= \Sigma^{++}_c \qquad S^{12}=\sqrt{\half} \Sigma^+_c \qquad
  S^{22} = \Sigma^0_c \cr
& \quad S^{13} = \sqrt{\half} {\Xi^+_c}' \qquad S^{23} = \sqrt{\half}
  {\Xi^0_c}' \cr
& \space\quad\qquad\qquad S^{33} = \Omega^0_c.  \cr}}
Their spin-$\threehalves$ counterparts are destroyed by corresponding
${S^*_\u}^{ij}(v)$ Rarita-Schwinger operators.  In the second case, the light
degrees of freedom form an antisymmetric $j_\ell =0$ combination which
transforms as a flavor antitriplet.  Coupling with the heavy quark then yields
$J^\P=\half^+$ baryons which we associate with the field $T_i(v)$.  When
$Q=c$, the individual components of $T_i$ are the singly charmed baryons
\eqn\antitripcomp{T_1=\Xi_c^0  \qquad T_2=-\Xi_c^+ \qquad T_3=\Lambda_c^+.}

	The complete spectrum of the first orbitally excited P-wave $Qqq$
baryons is quite complicated.  The lowest lying such hadrons correspond
to bound states that have one unit of orbital angular momentum inserted
between the heavy quark and light diquark pair.  In this case, spin statistics
constrain the light degrees of freedom to belong to either a $j_\ell = 1$
multiplet which transforms as a flavor antitriplet or else to $j_\ell=0$, $1$
or $2$ multiplets which transform as flavor sextets.  Nonrelativistic quark
model calculations indicate that the antitriplet multiplet is isolated and
lies significantly below all other P-wave states \Copley.  We will
therefore only incorporate this lightest $j_\ell=1$ multiplet into the
chiral Lagrangian. We assign the Dirac and Rarita-Schwinger operators
$R_i(v)$ and $R^*_{\u i}(v)$ to its $J^\P=\half^-$ and $J^\P=\threehalves^-$
states.  As we shall see, the two newly discovered $\Lambda_c^*$ baryons are
well described as the $I=0$ members of $R$ and $R^*_\u$.

	In the infinite heavy quark mass limit, it is useful to combine
together the degenerate $J=\half$ and $J=\threehalves$ members of the ground
state sextet and excited antitriplet multiplets into the baryon ``superfields''
\refs{\ChoI,\Georgi}
\eqn\superfields{\eqalign{
\CR_{\u i} &= \sqrt{\third} (\g_\u+v_\u) \gfive R_i + R^*_{\u i} \cr
\CS^{ij}_\u &= \sqrt{\third} (\g_\u+v_\u) \gfive S^{ij} + {S^*_\u}^{ij}.\cr}}
The $\CT_i$ superfield for the ground state antitriplet baryons is simply
identical to $T_i$.  The superfields transform under parity as
\eqn\parity{\eqalign{
\CR_\u ({\vec x},t) & \space {\buildrel \P \over \longrightarrow} \space \g_0
 \CR^\u(-{\vec x},t) \cr
\CS_\u ({\vec x},t) & \space {\buildrel \P \over \longrightarrow} \space - \g_0
 \CS^\u(-{\vec x},t)\cr
\CT ({\vec x},t) & \space {\buildrel \P \over \longrightarrow} \space \g_0
 \CT (-{\vec x},t)\cr}}
and obey the constraints
\eqn\constraints{
\eqalign{\proj \CR_\u &= \CR_\u \cr v^\u \CR_\u &= 0 \cr}
\qquad
\eqalign{\proj \CS_\u &= \CS_\u \cr v^\u \CS_\u &= 0 \cr}
\qquad
\proj \CT = \CT.}
These conditions ensure that $\CR_\u$ and $\CS_\u$ contain six degrees of
freedom while $\CT$ has two.  The degree of freedom count thus agrees with
the number of states that the superfields represent \ChoII.

	The constraints in \constraints\ also fix the shifts in the baryon
superfields induced by the reparameterization transformation
\eqn\paramshifts{\eqalign{v &\to v+\e/M \cr
			  k &\to k-\e \cr}}
where $v \cdot \e = 0$.  This change of variables leaves invariant the total
four-momentum $p=Mv+k$ of a heavy hadron and induces only an $O(1/M^2)$
correction to $v^2=1$.  The method for determining the induced shifts in the
baryon superfields is entirely analogous to that for their meson counterparts
which has previously been discussed in ref.~\ChoTrivedi.  So we only quote the
results here:
\eqn\supershifts{\eqalign{
\d \CR_\u &= {\slash{\e} \over 2M} \CR_\u -{\e^\v \CR_\v \over M} v_\u \cr
\d \CS_\u &= {\slash{\e} \over 2M} \CS_\u -{\e^\v \CS_\v \over M} v_\u \cr
\d \CT &= {\slash{\e} \over 2M} \CT.}}
The requirement that the effective theory remain invariant under the
transformations in \paramshifts\ and \supershifts\ forbids certain terms from
appearing in the chiral Lagrangian \LukeManohar.

	The heavy baryons in the $\CR_\u$, $\CS_\u$ and $\CT$ multiplets
can interact with one another via emission and absorption of light Goldstone
bosons.  The Goldstone bosons result from the spontaneous breaking of
$SU(3)_\L \times SU(3)_\R$ chiral symmetry down to its diagonal
$SU(3)_{\L+\R}$ flavor subgroup and appear in the pion octet
\eqn\pionoctet{\pib = \sum_{a=1}^8 \pi^a T^a = {1 \over \sqrt{2}}
\pmatrix{ \sqrt{\half} \pi^0 + \sqrt{\sixth} \eta & \pi^+ & K^+ \cr
\pi^- & - \sqrt{\half} \pi^0+\sqrt{\sixth}\eta & K^0 \cr
K^- & \bar{K}^0 & - \sqrt{\twothirds}\eta\cr}.}
It is convenient to arrange these fields into the exponentiated matrix
functions $\Sigma=e^{2i \pib/f}$ and $\xi=e^{i \pib/f}$ where the parameter
$f$ equals the pion decay constant $f_\pi=93 \MeV$ at lowest order.  The
matrix functions transform under the chiral symmetry group as
\eqn\Sigmafields{\eqalign{\Sigma &\to L \Sigma R^\dagger \cr
\xi &\to L \xi U^\dagger(x) = U(x) \xi R^\dagger \cr}}
where $L$ and $R$ represent global elements of $SU(3)_\L$ and $SU(3)_\R$ while
$U(x)$ acts like a local $SU(3)_{\L+\R}$ transformation.  We further define
the vector and axial vector fields
\eqn\pioncurrents{\eqalign{
{\bf V}^\u &= \half (\xi^\dagger \partial^\u \xi
  + \xi \partial^\u \xi^\dagger)
  = {1 \over 2f^2} [ \pib, \partial^\u \pib]-{1 \over 24 f^4} \Bigl[ \pib,
  \bigl[\pib, [\pib,\partial^\u \pib] \bigr] \Bigr] + O(\pib^6) \cr
{\bf A}^\u &= {i \over 2}(\xi^\dagger \partial^\u \xi - \xi \partial^\u
  \xi^\dagger) = -{1 \over f} \partial^\u \pib + {1 \over 6 f^3}
  \bigl[\pib,[\pib,\partial^\u \pib] \bigr] + O(\pib^5) \cr}}
which transform inhomogeneously and homogeneously under $SU(3)_{\L+\R}$
respectively:
\eqn\pioncurrtrans{\eqalign{
{\bf V}^\u &\to U {\bf V}^\u U^\dagger + U \partial^\u U^\dagger \cr
{\bf A}^\u &\to U {\bf A}^\u U^\dagger. \cr}}
The pions in \pionoctet\ derivatively couple to the baryon matter fields via
these vector and axial vector combinations.

	It is straightforward to construct the lowest order
effective Lagrangian which describes the low energy interactions
between the $Qqq$ baryons and Goldstone bosons.  One simply writes down all
possible terms that are Lorentz invariant, light chiral and heavy quark
spin symmetric, and parity even:
\eqn\Lzero{\eqalign{
\CL_v^{(0)} &= \sum_{Q=c,b} \Bigl\{
 \bar{\CR}_\u^i \bigl( -i v\cdot\CD + \Delta M_\CR \bigr) \CR^\u_i
+\bar{\CS}^\u_{ij} \bigl( -i v\cdot\CD + \Delta M_\CS \bigr) S^{ij}_\u
+\bar{\CT}^i i v \cdot \CD \, \CT_i \cr
& \qquad\qquad + i g_1 \, \varepsilon_{\u\v\o\l} \bar{\CS}^\u_{ik} \,
  v^\v (A^\o)^i_j (\CS^\l)^{jk}
+ i g_2 \, \varepsilon_{\u\v\o\l} \bar{\CR}^{\u i}
  v^\v (A^\o)^j_i (\CR^\l)_j \cr
& \qquad\qquad + h_1 \Bigl[ \e_{ijk} \bar{\CT}^i (A^\u)^j_l {\CS}^{kl}_\u
  + \e^{ijk} \bar{\CS}^\u_{kl} (A_\u)^l_j \CT_i \Bigr] \cr
& \qquad\qquad +h_2\Bigl[ \e_{ijk} \bar{\CR}^{\u i} \, v\cdot A^j_l \CS^{kl}_\u
  +\e^{ijk}\bar{\CS}^\u_{kl} \, v\cdot A^l_j \CR_{\u i} \Bigr] \Bigr\}. \cr}}
A few points about this zeroth order Lagrangian should be noted.  Firstly,
the common mass splitting between the excited and ground state antitriplet
multiplets is absorbed into the parameter $\Delta M_\CR=M_\CR-M_\CT$.
Similarly, $\Delta M_\CS = M_\CS-M_\CT$ represents the splitting
between the ground state sextet and antitriplet multiplets.
These parameters do not vanish in the zero or infinite heavy quark mass
limits and therefore appropriately reside within the leading order chiral
Lagrangian.  Secondly, the coupling constants $g_{1,2}$ and $h_{1,2}$ in
\Lzero\ are expected to be of order unity on general dimensional analysis
grounds \NDA.  However, their precise numerical values are {\it a priori}
unknown and must be fitted to experiment.  Finally, we observe that there
are no terms in \Lzero\ which mediate the single Goldstone boson transitions
$R^{(*)} \to T \pib$ and $T \to T \pib$.  Such processes violate heavy
quark spin symmetry and occur only at next-to-leading order in the
$1/\mQ$ expansion.

	The current experimental status of the baryons appearing in the
heavy hadron chiral Lagrangian is very uneven.   Data on strange charmed
baryons is in short supply, and several have not yet been discovered.
In contrast, a number of experiments within the past year have filled in most
of the nonstrange members of the antitriplet and sextet multiplets.  We will
therefore focus upon the zero strangeness baryons in the remainder of this
work.

	The energy levels of the observed $\Lambda_c^{(*)}$ and
$\Sigma_c^{(*)}$ states in $\CR_\u$, $\CS_\u$ and $\CT$ are illustrated in
\spectrum.  As indicated in the figure, we interpret the two recently observed
excited charmed baryons as the $I=0$ members of the $\CR_\u$ multiplet.  In
the absence of well-established names for these baryons, we adopt the
nomenclature convention of ref.~\Appel\ and denote the $J^\P=\half^-$ and
$J^\P=\threehalves^-$ states as $\Lambda_{c1}$ and
$\Lambda_{c1}^*$ respectively.  Averaging over the ARGUS,
CLEO and E687 values for their masses, we find that they lie
$308.0 \pm 2.0 \MeV$ and $341.4 \pm 0.4 \MeV$ above $\Lambda_c$.
The splitting between these two P-wave baryon masses is comparable in
magnitude to that between their P-wave meson analogues $D_1(2421 \MeV)$ and
$D_2(2465 \MeV)$.  We will keep track of this phenomenologically important
mass difference even though it represents an $O(1/\mc)$ effect.

	The splitting between $\Sigma_c^*$ and $\Lambda_c$ displayed in
\spectrum\ comes from another recent experimental result.  The SKAT group
claims to have observed the $J^\P = \threehalves^+$ $\Sigma_c^{* ++}$ baryon
for the first time in their bubble chamber experiment which uses a broad-band
neutrino beam \SKAT.  While their mass finding
$M_{\Sigma^*_c} = 2530 \pm 7 \MeV$ must be treated with caution until
independently confirmed by another group, we will adopt their reasonable
central value in our subsequent analysis.  Fortunately, none of our results
will sensitively depend upon the precise numerical value for the $\Sigma_c^*$
mass.

	Having set up the necessary machinery for studying the two new
$\Lambda_c^*$ baryons, we proceed to examine their strong and
radiative decay modes in the following two sections.

\newsec{Strong Decays of $\pmb{\Lambda_c^*}$}

	The strong decays of the newly discovered excited charmed baryons
are well-suited for Chiral Perturbation Theory analysis.  The relatively
small masses of $\Lambda_{c1}$ and $\Lambda_{c1}^*$ above $\Lambda_c$
kinematically restrict their strong decays to soft pion emission.  We
therefore expect the chiral Lagrangian derivative expansion to be well-behaved
for these new particles.  Moreover, isospin conservation forbids single pion
transitions between $\Lambda_{c1}^{(*)}$ and $\Lambda_c$.  The excited $I=0$
baryons must instead decay via an intermediate $I=1$ state down to the $I=0$
ground state.  The released energy $M_{\Lambda_{c1}^{(*)}} - M_{\Lambda_c}$
is thus shared by two
pions.~\foot{The analogous kinematics for excited P-wave mesons is much
less favorable.  For example, the splitting between the $D_2$ and $D$ mesons
is almost $600 \MeV$, and single pion transitions between these two states
are allowed.  The validity of lowest order Chiral Perturbation Theory in this
case is dubious at best.}

	Angular momentum and parity considerations require single pion
transitions between the $\CR_\u$ and $\CS_\u$ multiplets to go through
$L=0$ or $L=2$ partial waves.  The D-wave coupling arises from dimension-five
operators in the next-to-leading order chiral Lagrangian whose effects are
quite suppressed.  The S-wave coupling on the other hand is implemented by the
dimension-four term proportional to $h_2$ in \Lzero\ which links
$\Lambda_{c1}$ with $\Sigma_c$ and $\Lambda_{c1}^*$ with $\Sigma_c^*$.
The $h_2$ operator consequently mediates the barely allowed transition
$\Lambda_{c1} \to \Sigma_c \pi$ at the rate
\eqn\singlepionrate{
\Gamma(\Lambda_{c1} \to \Sigma_c \pi) = {h_2^2 \over 4 \pi f^2}
{M_{\Sigma_c} \over M_{\Lambda_{c1}}} (M_{\Lambda_{c1}}-M_{\Sigma_c})^2
\sqrt{(M_{\Lambda_{c1}} - M_{\Sigma_c})^2 - \mpi^2}.}
This process occurs so close to threshold that small isospin violating
mass differences between members of the pion and charmed Sigma baryon
multiplets cannot be ignored in the phase space factors of \singlepionrate.
Using the values
$M_{\Sigma_c^0}=2452.0 \MeV$, $M_{\Sigma_c^+}=2453.4 \MeV$ and
$M_{\Sigma_c^{++}}=2453.1 \MeV$ \Sigmamasses, we find the partial widths
\eqna\partialwidths
$$ \eqalignno{
\Gamma(\Lambda_{c1}^+ \to \Sigma_c^0 \pi^+) &= 3.3 h_2^2 \MeV &
\partialwidths a \cr
\Gamma(\Lambda_{c1}^+ \to \Sigma_c^+ \pi^0) &= 6.0 h_2^2 \MeV &
\partialwidths b \cr
\Gamma(\Lambda_{c1}^+ \to \Sigma_c^{++} \pi^-) &= 1.4 h_2^2 \MeV. &
\partialwidths c \cr} $$
The analogous single pion transitions between the $J=\threehalves$ baryons in
$\CR_\u$ and $\CS_\u$ are kinematically forbidden.

	Double pion decays of $\Lambda_{c1}$ and $\Lambda_{c1}^*$ down to
the $\Lambda_c$ ground state proceed at leading order via the two pole graphs
displayed in \polegraphs.  In order to obtain convergent decay rates from
these diagrams, we must take into account the nonzero widths
\eqn\widths{\eqalign{
\Gamma_{\Sigma_c} &= {h_1^2\over 12\pi f^2} {M_{\Lambda_c} \over M_{\Sigma_c}}
\bigl[(M_{\Sigma_c}-M_{\Lambda_c})^2-\mpi^2 \bigr]^{3/2} \simeq
 2.5 h_1^2 \MeV \cr
\Gamma_{\Sigma_c^*}  &= {h_1^2 \over 12 \pi f^2}
{M_{\Lambda_c} \over M_{\Sigma_c^*}}
\bigl[(M_{\Sigma_c^*}-M_{\Lambda_c})^2-\mpi^2 \bigr]^{3/2} \simeq
 24 h_1^2 \MeV \cr}}
of the intermediate $\Sigma_c$ and $\Sigma_c^*$ resonances.  Their
propagators thus appear as
\eqn\propagators{\eqalign{
D_{\Sigma_c} &=
  {i \over v\cdot k-(M_{\Sigma_c}-M_{\Lambda_c})+ i\Gamma_{\Sigma_c}/2}
  \, \Lambda_+ \cr
D^{\u\v}_{\Sigma^*_c} &=
  {i \over v\cdot k-(M_{\Sigma^*_c}-M_{\Lambda_c})+i\Gamma_{\Sigma_c^*}/2} \,
  \Lambda^{\u\v}_+ \cr}}
where $\Lambda_+ = (1 + \slash{v})/2$ and $\Lambda_+^{\u\v} = \bigl[
-g^{\u\v}+v^\u v^\v + \third(\g^\u+v^\u)(\g^\v-v^\v) \bigr] \Lambda_+$
denote spin-$\half$ and spin-$\threehalves$ projection operators respectively.
We must also include a symmetry factor of $1/2$ in the angular integration
over the pions' momenta to avoid double counting the two identical
bosons in the final state.  A straightforward computation then yields
the dimensionless differential decay rate
\eqn\diffdecayrate{\eqalign{
& {d\Gamma \bigl(\Lambda_{c1}^{(*)}\to\Lambda_c \pi^a \pi^b \bigr) \over dE_1}
= {\d^{ab} \over 192 \pi^3} \Bigl({h_1 h_2 \over f^2} \Bigr)^2
{M_{\Lambda_c} \over M_{\Lambda_{c1}^{(*)}}} \sqrroot \cr
&\qquad \times \biggl[ { E_1^2(E_2^2-\mpi^2) \over
\bigl(M_{\Lambda_{c1}^{(*)}}-M_{\Sigma_c^{(*)}}-E_1\bigr)^2
+\Gamma_{\Sigma_c^{(*)}}^2/4}
 + {(E_1^2-\mpi^2) E_2^2 \over
\bigl(M_{\Lambda_{c1}^{(*)}}-M_{\Sigma_c^{(*)}}-E_2 \bigr)^2
+\Gamma_{\Sigma_c^{(*)}}^2/4} \biggr]\cr}}
expressed in terms of the two pion energies $E_1$ and
$E_2 = M_{\Lambda_{c1}^{(*)}}-M_{\Lambda_c}-E_1$ measured in the decaying
body's rest frame.
\foot{In the infinite charm mass limit, the recoiling $\Lambda_c$
baryon carries off momentum but no kinetic energy.  The two pions thus
share all of the energy released by the decaying $\Lambda_{c1}^{(*)}$.  This
situation is similar to bouncing a ball off the earth.  The earth must
recoil to conserve momentum, but the ball bounces back with practically all
its original kinetic energy \bounce.}
Integrating over $E_1$, we obtain the total rate
\eqn\totaldecayrate{
\Gamma \bigl(\Lambda_{c1}^{(*)} \to \Lambda_c \pi^a \pi^b \bigr) =
{h_2^2 \d^{ab} \over 8\pi^2 f^2}
{M_{\Sigma_c^{(*)}} \over M_{\Lambda_{c1}^{(*)}}}
{I\over\biggl[(M_{\Sigma_c^{(*)}}-M_{\Lambda_c})^2-\mpi^2\biggr]^{3/2}}}

where
\eqn\integral{\eqalign{
I &= {\Gamma_{\Sigma_c^{(*)}} \over 2}
\int^{M_{\Lambda_{c1}^{(*)}}-M_{\Lambda_c}-\mpi}_{\mpi} dE_1 \sqrroot \cr
&\qquad \times \biggl[
{E_1^2(E_2^2-\mpi^2) \over
\Bigl(M_{\Lambda_{c1}^{(*)}}-M_{\Sigma_c^{(*)}}-E_1 \Bigr)^2
+ \Gamma_{\Sigma_c^{(*)}}^2/4} +
{(E_1^2-\mpi^2) E_2^2 \over
\Bigl(M_{\Lambda_{c1}^{(*)}}-M_{\Sigma_c^{(*)}}-E_2 \Bigr)^2
+ \Gamma_{\Sigma_c^{(*)}}^2/4}
\biggr]. \cr}}

	Since we do not know the values of $h_1$ and $h_2$, we
cannot extract precise quantitative predictions from eqns.~\diffdecayrate\
$\dash$ \integral.  However, these formulas do provide useful qualitative
insight into the P-wave baryons' strong decays.  In \diffrate, we plot
$h_2^{-2} d\Gamma(\Lambda_{c1}^{(*)} \to \Lambda_c \pi^0 \pi^0)/dE_1$ versus
$E_1$ with $h_1$ set equal to unity.  As can clearly be seen
in the figure, $\Lambda_{c1} \to \Lambda_c\pi^0 \pi^0$ is
dominated by the pole regions where the intermediate $\Sigma_c^+$ state is
very close to being on-shell.  Its integrated rate is thus well approximated
by the single $\pi^0$ partial width in \partialwidths{b}.  The rate in the
charged pion channel is similarly well approximated by the sum of the
widths in \partialwidths{a}\ and \partialwidths{c}.  Indeed, evaluating the
phase space integral in \integral\ using the narrow width approximation
\eqn\narrowwidth{{\Gamma_{\Sigma_c}/2 \over
(M_{\Lambda_{c1}}-M_{\Sigma_c}-E)^2+\bigl(\Gamma_{\Sigma_c}/2\bigr)^2} \simeq
\pi \d(M_{\Lambda_{c1}}-M_{\Sigma_c}-E),}
we simply recover eqn.~\singlepionrate\ for
$\Gamma(\Lambda_{c1} \to \Sigma_c \pi)$ which is independent of coupling
constant $h_1$.

	Nonresonant contributions generate a slight dependence
of $\Gamma(\Lambda_{c1} \to \Lambda_c \pi^a \pi^b)$ upon $h_1$ as shown in
\twopirate.  But the decay of the $J^\P = \half^-$ state may essentially be
viewed as the two step process $\Lambda_{c1} \to \Sigma_c \pi$ followed by
$\Sigma_c \to \Lambda_c \pi$.  In contrast, the double pion decay of
$\Lambda_{c1}^*$ cannot be regarded as a sequential transition.  The virtual
$\Sigma_c^*$ intermediate state is very much off-shell and produces no large
resonant contribution to $\Lambda_{c1}^* \to \Lambda_c \pi \pi$.  As a result,
the strong interaction partial width of the $J^\P = \threehalves^-$ state is
more than an order of magnitude smaller than that of its $J^\P = \half^-$
partner.

	As advertised in the Introduction, these qualitative findings on the
excited charm baryon decay modes are in basic accord with the recent CLEO
results reported in refs.~\CLEOI\ and \CLEOII. They thus bolster one's
confidence in the interpretation of the two new states as
$\Lambda_c^*$ baryons.  To make further progress however, we need width
information to pin down the
values of the coupling constants in chiral Lagrangian \Lzero.  ARGUS has set
a $90\%$ CL upper bound of $3.2 \MeV$ on the width of $\Lambda_{c1}^*$ \ARGUS.
Unfortunately, this limit places only a weak constraint on the allowed
parameter space in the $h_1 \dash h_2$ plane.  As \twopirate\ demonstrates,
the true natural width of $\Lambda_{c1}^*$ is most likely too narrow to be
resolved by current experimental detectors.  On the other hand, there is a
much better chance that the $\Lambda_{c1}$ resonance is wide enough to be
measured.  In the $I=1$ sector, the width resolving prospects for the
$J^\P = \half^+$ and $J^\P = \threehalves^+$ members of the $\CS_\u$ doublet
are just opposite those for the $J^\P = \half^-$ and $J^\P = \threehalves^-$
members of $\CR_\u$.  We are therefore hopeful that experimentalists will
be able to fix some of the free parameters in the heavy baryon chiral
Lagrangian in the near future.

\newsec{Electromagnetic Decays of $\pmb{\Lambda_c^*}$}

	The only decay modes of the two new $\Lambda_{c1}^{(*)}$
baryons that have so far been experimentally observed are their double pion
transitions to $\Lambda_c$.  But as shown in \spectrum, these P-wave
hadrons can also de-excite down to the ground state via single photon
emission.  Unlike the strong interaction processes, the radiative channels
are not severely phase space suppressed.  Moreover, they produce two rather
than three bodies in the final state.  So the inherently weaker strength of
the electromagnetic transitions could be offset by their more favorable
kinematics.  We explore such a possibility in this section.

	Electromagnetic interactions may be incorporated into Heavy Hadron
Chiral Perturbation Theory by gauging a $U(1)_\EM$ subgroup of the global
$SU(3)_\L \times SU(3)_\R$ chiral symmetry group.  All derivatives appearing
in the velocity dependent effective Lagrangian are then promoted to covariant
derivatives with respect to electromagnetism.  The leading dimension-four
operators in \Lzero\ cannot contribute to S-matrix elements between
states containing real photons.  So to study heavy meson and baryon radiative
transitions, one must include a number of dimension-five operators into the
chiral Lagrangian.  In refs.~\refs{\ChoGeorgi{--}\YanII}, the $M1$
transitions between ground state hadrons containing a single heavy quark were
analyzed.  We now extend this earlier work to investigate the $E1$ decays
of the new $\Lambda_{c1}^{(*)}$ baryons.

	In the low energy theory, short wavelength photons with energies
greater than the chiral symmetry breaking scale are integrated out and only
long wavelength modes are retained.  At lowest order in the $1/\mQ$ expansion,
photons couple to just the light brown muck inside $Qqq$ baryons leaving the
spins of their heavy quark constituents unaltered.  Such couplings
generate the following contributions to the effective Lagrangian:
\eqn\LEM{\eqalign{
\CL_v^\EM &= \sum_{Q=c,b} {e(\LX) \over \LX} \Bigl\{
  i \cR \bar{\CR}_\u^j \CQ^i_j \CR_{\v i}  F^{\u\v}
+ i \cS\bar{\CS}_{\v ij} \bigl(\CQ^i_k \CS^{kj}_\u + \CQ^j_k \CS^{ik}_\u \bigr)
  F^{\u\v} \cr
& \quad + \cRS \Bigl[ \e_{ijk} \bar{\CR}_\u^i \CQ^j_l \CS^{kl}_\v
  + \e^{ijk} \bar{\CS}_{\v,kl} \CQ^l_j \CR_{\u i} \Bigr] {\tilde F}^{\u\v}
  + \cRT \Bigl[ \bar{\CT}^j \CQ^i_j \CR^\u_i
  + \bar{\CR}^{\u i} \CQ^j_i \CT_j \Bigr] v^\v F_{\u\v} \cr
& \quad + \cST \Bigl[ \e_{ijk} \bar{\CT}^i \CQ^j_l \CS^{kl}_\v
  + \e^{ijk} \bar{\CS}_{\v,kl} \CQ^l_j \CT_i \Bigr]
   v_\u {\tilde F}^{\u\v} \Bigr\}. \cr}}
Here $F^{\u\v}$ and ${\tilde F}^{\u\v}$ are the electromagnetic field strength
tensor and its dual, and ${\pmb\CQ} = \half(\xi Q_\EM \xi^\dagger +
\xi^\dagger Q_\EM \xi)$ where
\eqn\Qcharge{Q_\EM = \pmatrix{\CQ_u && \cr & \CQ_d & \cr && \CQ_s \cr}
       = \pmatrix{\twothirds && \cr & -\third & \cr && - \third\cr}}
denotes the light quark electric charge matrix.  The transformation rule
${\pmb\CQ} \to U {\pmb\CQ} U^\dagger$ for this spurion field renders the
terms in \LEM\ chiral symmetric \YanIII.

	The interactions between low energy photons and light brown muck take
place at a long distance scale $\LX$.   In Chiral Perturbation
Theory, this parameter represents the chiral symmetry breaking scale whose
numerical value is approximately $1000 \MeV$.  The phenomenologically
successful Nonrelativistic Quark Model suggests however that a more
appropriate value for $\LX$ in $\CL_v^\EM$ is a typical constituent quark
mass of $300 \MeV$.  We cannot really distinguish between these two numbers
at the level of naive dimensional analysis as they differ by only a factor
of three.  But since $\LX$ enters quadratically into radiative decay rates,
it is important to minimize its uncertainty as much as possible.  So we will
compromise and take $\LX$ to be the geometric mean between the CPT and NRQM
values:
\eqn\LXvalue{\LX^2 = (300 \MeV)(1000 \MeV) = (547.7 \MeV)^2.}
Hopefully this guess for $\LX^2$ does not lie more than a factor of three
away from the true value.

	The terms in \LEM\ proportional to $\cR$, $\cS$ and $\cST$ mediate
the $M1$ radiative transitions $R^* \to R \g$, $S^* \to S \g$ and
$S^{(*)} \to T \g$.  The $\cRS$ and $\cRT$ operators on the other hand
generate $E1$ decays $R^{(*)} \to S^{(*)} \g$ and $R^{(*)} \to T \g$.
After extracting the $\Lambda_{c1}^{(*)}$ components from these
interactions, we find the following radiative partial widths:
\eqn\photratesI{
\Gamma(\Lambda_{c1}^* \to \Lambda_{c1} \g) = {4 \cR^2 \over 81}
{\aEM(\LX) \over \LX^2} {M_{\Lambda_{c1}} \over M_{\Lambda_{c1}^*}}
\biggl({ M_{\Lambda_{c1}^*}^2 - M_{\Lambda_{c1}}^2 \over
2 M_{\Lambda_{c1}^*}} \biggr)^3 = 4.36 \times 10^{-5} \, \cR^2 \MeV}
\eqna\photratesII
$$ \eqalignno{
\Gamma(\Lambda_{c1} \to \Sigma_c \g) &= {8 \cRS^2 \over 9}
{\aEM(\LX) \over \LX^2} {M_{\Sigma_c} \over M_{\Lambda_{c1}}}
\biggl({ M_{\Lambda_{c1}}^2 - M_{\Sigma_c}^2 \over
2 M_{\Lambda_{c1}}} \biggr)^3 = 0.052 \, \cRS^2 \MeV & \photratesII a\cr
\Gamma(\Lambda_{c1} \to \Sigma_c^* \g) &= {4 \cRS^2 \over 9}
{\aEM(\LX) \over \LX^2} {M_{\Sigma_c^*} \over M_{\Lambda_{c1}}}
\biggl({ M_{\Lambda_{c1}}^2 - M_{\Sigma_c^*}^2 \over
2 M_{\Lambda_{c1}}} \biggr)^3 = 0.003 \, \cRS^2 \MeV & \photratesII b\cr
\Gamma(\Lambda_{c1}^* \to \Sigma_c \g) &= {2 \cRS^2 \over 9}
{\aEM(\LX) \over \LX^2} {M_{\Sigma_c} \over M_{\Lambda_{c1}^*}}
\biggl({ M_{\Lambda_{c1}^*}^2 - M_{\Sigma_c}^2 \over
2 M_{\Lambda_{c1}^*}} \biggr)^3 = 0.024 \, \cRS^2 \MeV & \photratesII c\cr
\Gamma(\Lambda_{c1}^* \to \Sigma_c^* \g) &= {10 \cRS^2 \over 9}
{\aEM(\LX) \over \LX^2} {M_{\Sigma_c^*} \over M_{\Lambda_{c1}^*}}
\biggl({ M_{\Lambda_{c1}^*}^2 - M_{\Sigma_c^*}^2 \over
2 M_{\Lambda_{c1}^*}} \biggr)^3 = 0.023 \, \cRS^2 \MeV & \photratesII d\cr} $$
\eqna\photratesIII
$$ \eqalignno{
\Gamma(\Lambda_{c1} \to \Lambda_c \g) &= {4 \cRT^2 \over 27}
{\aEM(\LX) \over \LX^2} {M_{\Lambda_c} \over M_{\Lambda_{c1}}}
\biggl({ M_{\Lambda_{c1}}^2 - M_{\Lambda_c}^2 \over
2 M_{\Lambda_{c1}}} \biggr)^3 = 0.103 \,\cRT^2 \MeV & \photratesIII a \cr
\Gamma(\Lambda_{c1}^* \to \Lambda_c \g) &= {4 \cRT^2 \over 27}
{\aEM(\LX) \over \LX^2} {M_{\Lambda_c} \over M_{\Lambda_{c1}^*}}
\biggl({ M_{\Lambda_{c1}^*}^2 - M_{\Lambda_c}^2 \over
2 M_{\Lambda_{c1}^*} } \biggr)^3 = 0.078 \,\cRT^2 \MeV. & \photratesIII b
\cr}$$
As required heavy quark spin symmetry, the sum of the widths in
\photratesII{a,b}\ equals the sum of those in \photratesII{c,d}\ in the
infinite charm mass limit \IsgurWise.  Similarly, the rates in
\photratesIII{a}\ and \photratesIII{b}\ become degenerate when
$\mc \to \infty$.

	The leading order results in eqns.~\photratesI\ $\dash$
\photratesIII{}\ cannot be trusted to provide much more than order of
magnitude estimates for the $\Lambda_{c1}^{(*)}$ radiative decay rates.
Yet comparing these electromagnetic partial widths with their strong
interaction counterparts, we can draw some general qualitative conclusions.
Firstly, we expect on the basis of naive dimensional analysis that the
unknown $\cR$, $\cRS$ and $\cRT$ couplings are of order unity.
The numerical partial width estimates suggest that some of the
electromagnetic branching fractions might be measurable.  Referring to
\twopirate, we see that the two pion decay mode of $\Lambda_{c1}$ dominates
over its radiative channels.  The electromagnetic branching fraction for the
$J^\P=\half^-$ state is thus most likely less than a few percent.  On the
other hand, since the double pion width of
$J^\P=\threehalves^-$ $\Lambda_{c1}^*$ is much more narrow,
$\Gamma(\Lambda_{c1}^* \to \Lambda_c \g) /
\Gamma(\Lambda_{c1}^* \to \Lambda_c \pi \pi)$ could be sizable and perhaps
greater than unity.  Finally, we note that the radiative mode
$\Lambda_{c1}^* \to \Sigma_c^* \g$ may provide a means for detecting the
$\Sigma_c^*$ baryon.  The branching fraction for this process is small
but not negligible.  A search for this transition could therefore
yield evidence for the elusive $I=1$, $J^\P = \threehalves^+$ state.

\newsec{Conclusion}

	The basic interpretation of the two new excited charm baryons
as $I=0$ members of an excited P-wave doublet holds together remarkably well.
Since the splitting between $\Lambda_{c1}^{(*)}$ and $\Lambda_c$
is relatively small, these excited hadrons are well suited for incorporation
into Heavy Hadron Chiral Perturbation Theory.  Many experimental and
theoretical details clearly remain to be filled into the picture
which we have outlined here.  In particular, width and branching ratio
information are needed to fix the several new parameters that enter into the
excited baryon sector.

	A number of extensions of this work could be pursued in the future.
For example, the primary decay modes of the $\Xi_{c1}^{0(*)}$ and
$\Xi_{c1}^{+(*)}$ partners of $\Lambda_{c1}^{(*)}$ ought to be analyzed.
As we have seen, there is no leading order term which links any of the states
in the $\CR_\u$ antitriplet superfield with members of the ground state $\CT$
multiplet.  So if kinematically allowed, single kaon decays of these
P-wave strange charmed baryons down to $\Lambda_c$ are suppressed by $1/\mc$.
A theoretical study of the dominant $\Xi_{c1}^{(*)}$ transitions would
help guide an experimental search for these states.  Alternatively, one might
consider including other excited P-wave $Qqq$ baryons into the heavy chiral
Lagrangian.  There are many such states waiting to be discovered.

	In short, excited heavy baryon physics is a subject in which we
may look forward to experimental and theoretical progress in the near future.

\bigskip\bigskip
\centerline{\bf Acknowledgments}
\bigskip

	Helpful discussions with Sandip Trivedi and Mark Wise are gratefully
acknowledged.

\bigskip

\listrefs
\listfigs
\bye